\documentclass[twocolumn,pre,superscriptaddress]{revtex4}

\usepackage{graphicx}
\usepackage{dcolumn}
\usepackage{bm}

\newcommand{\avg}[1]{\left\langle{#1}\right\rangle}
\newcommand{\eref}[1]{(\ref{#1})}
\newcommand{\figref}[1]{Fig.~\ref{#1}}
\newcommand{\tabref}[1]{Table~\ref{#1}}
\newcommand{\etal}{{\it{}et~al.}}

\begin{document}

\title{A Gap in the Community-Size Distribution of a Large-Scale
  Social Networking Site}

\author{Kikuo Yuta}
\email[e-mail: ]{yuta@crev.jp}
\affiliation{NiCT/ATR CIS, Applied Network Science Laboratory, Kyoto
  619-0288, Japan}
\author{Naoaki Ono}
\email[e-mail: ]{nono@bio.eng.osaka-u.ac.jp}
\affiliation{Department of Bioinformatics Engineering,
  Graduate School of Information Science and Technology,
  Osaka University, 2-1 Yamadaoka, Suita, Osaka 565-0871, Japan}
\author{Yoshi Fujiwara}
\email[e-mail: ]{yfujiwar@atr.jp}
\affiliation{NiCT/ATR CIS, Applied Network Science Laboratory, Kyoto
  619-0288, Japan}


\begin{abstract}
  Social networking sites (SNS) have recently used by millions of
  people all over the world. An SNS is a society on the
  Internet, where people communicate and foster friendship with each
  other. We examine a nation-wide SNS (more than six million users at
  present), mutually acknowledged friendship network with third
  million people and nearly two million links. By employing a
  community-extracting method developed by Newman and others,
  we found that there exists a range of community-sizes in which only
  few communities are detected. This novel feature cannot be explained
  by previous growth models of networks.  We present a simple model
  with two processes of acquaintance, connecting nearest neighbors and
  random linkage. We show that the model can explain the gap in the
  community-size distribution as well as other statistical properties
  including long-tail degree distribution, high transitivity, its
  correlation with degree, and degree-degree correlation.
  The model can estimate how the two processes, which are ubiquitous in
  many social networks, are working with relative frequencies in the
  SNS as well as other societies.
\end{abstract}


\maketitle

\section{Introduction}\label{sec:intro}

The last few years witnessed the emergence of a new channel of human
communication in the World Wide Web. This is called {\it social
 networking sites\/} (SNS). An SNS provides an arena on
the Internet, where millions of people are creating personal pages,
featuring profiles, photos, music, movies, daily records etc., and
at a same time, they are watching activities of others and
occasionally responding to some of them.
People frequently have communication with each other
by sending messages, during chats on same subjects, in on-line communities or
groups of people with similar interests, and thus grow up friendship.

An early example is {\tt Friendster}\footnote{%
  {\tt http://www.friendster.com}%
} for which a million people, in
a single quarter of 2003, had registered. {\tt MySpace}\footnote{%
  {\tt http://www.myspace.com}%
} attracted more than a hundred million people in the end of 2006, which
ranks fifth among all the Internet access to every WWW sites. Other
sites include {\tt orkut}\footnote{%
  {\tt https://www.orkut.com}%
} with 36 million accounts, 65\% in Brazil,
{\tt Cyworld}\footnote{%
  {\tt http://www.cyworld.com}%
}, 18 million mostly in Korea, and {\tt Facebook}\footnote{%
  {\tt http://www.facebook.com}%
}, more than 13 million, 85\% students in USA (the numbers are so
recorded at the time of writing). We believe that the present reader
has experience in one SNS or more, and that it is not disputable that
these sites provide {\it societies\/} on the Internet, which form giant
human networks.

This fact that recently an increasing amount of social interactions
are recorded electronically can boost the understanding of the
structural formation of human networks in a society-wide scale, which
had never been accessible. Indeed, traditional social network studies (see
\cite{scott2000sna} for review) usually carry out collection of data
by querying people using questionnaires or interviews. Such methods
have been limiting the size of the network under study. Additionally,
survey data are often relying on individual's memory even to list up
friends. Researchers can now access to social networks of much larger
scale and of different nature. See the studies on e-mail
\cite{ebel2002sft,smith2002ims,tyler2003esa,kossinets2006eae}, phone calls
\cite{aielo2000rgm,onnela2006ssm}, for example, in this new direction.

There are some works on social networks on the Internet. Sociologists
have attempted to measure how the Internet and the Web services have
effect on real-life social interactions. Such effect is present in
occasions of ``off-line'' social events and ``on-line'' communities as
integrated patterns of social life (see Wellman's viewpoint
\cite{wellman2002eie} on this matter). Holme~\etal~\cite{holme2004ste}
investigated a dating site. It should be mentioned that a dating site
has different characteristics in the network structure, because the
incentives of participants in forming ties are relatively limited.
Actually, clustering coefficients are much lower than those in many
SNS. Adamic~\etal~\cite{adamic2003snc} studied a social networking
site at a university, which includes analysis of friendship, called
buddy, in relation to the attributes and personalities of the users.
Backstrom~\etal~\cite{backstrom2006gfl} investigated group formation
in an SNS, {\tt LiveJournal}\footnote{%
  {\tt http://www.livejournal.com}%
}, and a dataset of academic collaboration.
They focus on how on-line communities and interaction therein affect
group formation and network structure. It is remarked that in this
paper, we shall reserve the word, ``community'', to mean a tightly-knit
group of people in a linkage property, and distinguish it from
``on-line community'', an on-line group of people who have similar
interests but are not necessarily linked with each other.
See also the recent work \cite{golder2006ros} on messages exchanged by
users in {\tt Facebook}.

In this paper, we study a friendship network recorded at the largest
SNS in a country, which comprises more than third million people and
nearly two million links. Each link is a mutually acknowledged
friendship. Our main concern here is the community structure in the
network --- how people cluster into tightly-knit groups with
relatively high density, and how bunches of these groups are embedded
in the entire network. To uncover the structure of the giant network
of people, we employ a community-extracting method
\cite{newman2004fad,clauset2004fcs}.

In Section~\ref{sec:sns}, we describe the SNS, activities of users in
it, and the definition of link, namely friendship in the network.  In
Section~\ref{sec:stat}, we examine the structure of the friendship
network. In particular, we show that the network has a scale-free
degree distribution in its tail, high transitivity, and positive
degree-degree correlation, as observed in many social networks. In
Section~\ref{sec:commun}, however, we found a novel feature in the
distribution of community-sizes that there is a gap in the
community-sizes such that only few communities are extracted. In
Section~\ref{sec:cnnr}, we propose a simple model with connecting
nearest neighbors and random linkage, and show that it can explain the
gap as well as the other statistical properties.

\section{Social networking site and dataset}\label{sec:sns}

The largest SNS in Japan, as of December 2006, is
{\tt mixi}\footnote{{\tt http://mixi.jp}}, which
had started with a small group in March 2004 and has been rapidly
growing, a million accounts in August 2005, 6.6 million in November
2006. Our dataset, as of March 2005, is consisted of third million
accounts, nearly two million links of on-line friendship,
and about a million on-line communities. Individuals in all these data are encrypted
for privacy protection. The
number of users is roughly 10\% of all the domestic people who have
access to the Internet, from teenagers to adults, equally males and
females, including workers and non-workers. Mobile-phone users have
access at any location and time. At this epoch, the number
was growing as a power function of physical time with exponent 2 to
2.6, which implies that the rate of growth was proportional to the
user-number to the power 0.5 to 0.6.  Since the start, it has been
reported that about 70\% of the accounts visit the sites at least once
in three days week by week. Indeed, according to a survey\footnote{%
  Alexa Web Information Service: {\tt http://www.alexa.com}%
}, the {\tt mixi} is the third
most active SNS ({\tt MySpace} and {\tt orkut} are the top two) in
terms of access from users, matching {\tt Facebook} at activity.

Activities of the users are summarized as follows. A new person
participates in the site, provided that an already registered user
invites him or her who accepts the invitation. Otherwise the site is
not public to the Internet and is accessible only for the registered users.
This policy of publicity, which is taken by other SNS such as
{\tt orkut}, endows the site with a feature differing from blogs and
bulletin board systems in the WWW. While some SNS have different
policies about publicity, it is said that the users feel less fear and
anxiety about personal abuse and, actually, many users are observed to
name themselves as they do in real life, rather than anonymously.
This is presumably due to the invitation scheme, being invited by a
person, an acquaintance, to find oneself within many acquaintances.
Many people consider that this is a less uneasy environment to start
with.

After the registration, the users make their own profiles, write
diaries with varying frequencies, to which others make recommendations
and comments. Like other SNS, they are able to see logs of visitors
and to send and receive messages to anyone.
The profiles and diaries are selectively public either to
friends, to friends of friends, or to all in the SNS.

On-line communities are another design for promoting communication,
each with participants having shared interests and chats on same
subjects. A new on-line community is launched by an arbitrary user as
administrator, who sets its publicity either to participants or to the
entire SNS.  One can search particular persons and on-line communities
in the whole site by keywords and classified categories.

Through exchange of multiple information, from diaries to on-line
communities, one gets to know who has similar interests as his or
hers, and eventually become {\it friends\/} by mutual acknowledgement,
which is done by sending messages. This is the links and friendship
network which we study in this paper. One's friends are listed in
thumbnails at the top page of the user. Note that the devices of
diaries, footprints, lists of friends and on-line communities foster
growth of friendship collectively and in different ways. Even if a new
comer starts with a single link, he or she will quickly find
acquaintances at one or two steps in friends of friends, then
sometimes gets acquainted with more people noticed from footprints
or by search deep in the site.

The number of user accounts and links of friendships are respectively,
363,819 and 1,906,878, in our dataset. In average, one has about ten
friends. We shall examine more statistical properties in the next
section.

\section{Structure of friendship network}\label{sec:stat}

\begin{figure*}[htbp]
  \includegraphics[width=0.8\textwidth]{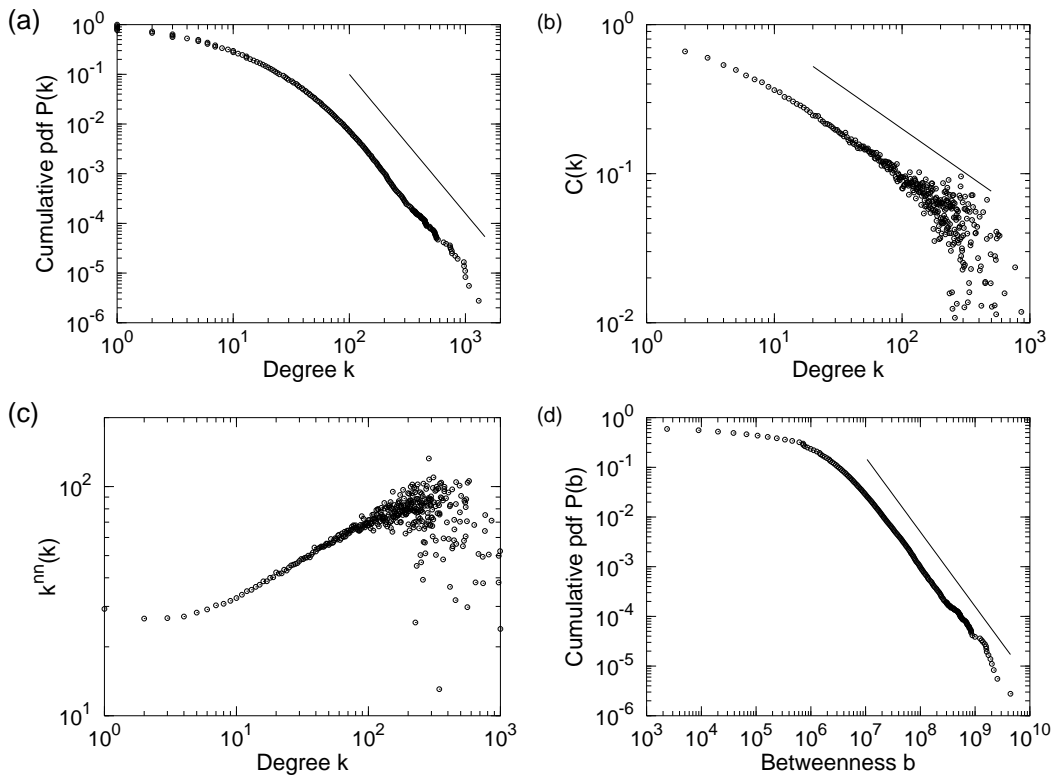}
  \caption{%
    For the largest connected component ($N=360,802$, $M=1,904,641$)
    of the friendship network, shown are %
    (a)~the cumulative degree distribution $P(k)$ for degree $k$, %
    (b)~the local clustering coefficient $C(k)$, %
    (c)~the nearest neighbor degree distribution $k^{\text{nn}}(k)$, %
    and (d)~the cumulative betweenness distribution $P(b)$. %
    The lines in (a), (b) and (c) are, respectively,
    $P(k)\propto k^{-1.8}$, $C(k)\propto k^{-0.6}$ and
    $P(b)\propto b^{-1.5}$.
  }
  \label{fig:stat}
\end{figure*}

\subsection*{Component structure and shortest paths}

People can be disconnected, because links are possible to
be lost by unregistration of users or by refusal of the corresponding
friendship. However, we found that most people are connected with each other. The
largest connected component, in fact, contains 360,802 people, 99.2\%
of all the users. The rest is composed of 1,213 disconnected
components, most of which are tiny groups each of a few people. We
examine the largest component in the following. Denoting the numbers
of nodes and links by $N$ and $M$ respectively, $N=360,802$ and
$M=1,904,641$. We use the words, participant and vertex,
interchangeably below.

Shortest-path lengths averaged over all pairs of vertices is given by
$\bar{d}=5.53$. The longest shortest-path has length
$d_{\text{max}}=22$, called diameter of the network. 

\subsection*{Degree distribution}

The number of links, or degree, have a long-tail distribution.
The degree distribution is denoted by $p_k$, i.e. the fraction of
vertices in the network with degree $k$. Cumulative degree
distribution is given by $P(k)=\sum_{k'=k}^\infty p_{k'}$. We plot
$P(k)$ in \figref{fig:stat}~(a).

The maximum degree is $k_{\text{max}}=1,301$. There are a small number
of hubs, about 100 people with links exceeding 300, even 3 persons
with degree 1000 or more. The time corresponding to the acquisition of
the data coincides when the site forbids participants to create more
than 1,000 links per each. This rule, however, did not essentially impose
a threshold in the degree studied here, as we have checked in
historical information of the site. On the other hand, 83,525 people
(23\%) have a single link, mostly new comers linked only to those who
invited; 182,125 people (50\%) have less than 5 links.

The first two moments of degree are
\begin{eqnarray}
  && \avg{k}=2M/N=10.56\ , \label{k1}\\
  && \avg{k^2}=593.4\ .    \label{k2}
\end{eqnarray}
The tail of degree distribution follows a power-law $p_k\propto
k^{-\alpha}$. The exponent was estimated in the region of $k$ greater
than 60 by the conventional mean-square-error in logarithmic
variables, and is given by $\alpha\sim1.80$, although the exponent
here, and the other ones given below, should be understood simply as
rough estimates.

\subsection*{Transitivity}

In many social networks, the friend of one's friend is quite likely
also to be the one's friend. Transitivity means how high the number of
triangles is present in the network (see the review \cite{newman2003sfc}). Global
clustering coefficient is defined by
\[ C_{\text g}=
\frac{3\times\text{number of triangles}}{\text{number of connected
    triples}} \ , \]
where a connected triple means a pair of vertices that are connected
to another node. $C_{\text g}$ is the mean probability that two
persons who have a common friend are also friends of each other. Our
dataset gives the value
\begin{equation}
  C_{\text g}=0.120=12\%\ .
  \label{Cg}
\end{equation}

To compare this with a class of random graphs which have the same size and
degree distribution, one can use the expected value of global
clustering coefficient given by \cite{newman2002rgm}
\begin{equation}
  C_{\text g}=\frac{\avg{k}}{N}\,\left[
    \frac{\avg{k^2}-\avg{k}}{\avg{k}^2}\right]^2
  \ .
  \label{random_Cg}
\end{equation}
Putting \eref{k1} and \eref{k2} into \eref{random_Cg} gives $C_{\text
 g}=8.0\times10^{-4}$, or 0.08\%. (For the class of Poisson random
graphs, \eref{random_Cg} reduces to $C_g=\avg{k}/N$, which is
$2.9\times10^{-5}$.) The observed value \eref{Cg} clearly shows strong
cliquishness in the local structure. People choose new acquaintances
who are friends of friends, well known as triadic closure.

Local clustering coefficient is a related and distinct measure of
cliquishness. For each vertex $i$, define
\[
  C_i=\frac{\text{number of triangles connected to }i}{\text{number of
  triples centered on }i}
  \ .
\]
The denominator is equal to $k_i(k_i-1)/2$ for the degree $k_i$ of the
vertex $i$. For $k_i=0$ and $1$, $C_i=0$ by convention. The averaged
clustering coefficient is then defined by $C=\sum_i C_i/N$. Our
dataset gives the value, $C=0.330$.

Local clustering coefficient $C_i$ has a strong dependence on the
degree $k_i$. To quantify it, one usually defines
\[ C(k)=\left.\avg{C_i}\right|_{k_i=k}\ . \]
In \figref{fig:stat}~(b), we plot the correlation between degree $k$
and $C(k)$.

We observe that $C(k)$ decreases as $k^{-0.6}$ for the range
$10\lesssim k\lesssim 200$. This differs from many other networks, where
$C(k)\sim k^{-1}$ gives a fit as reported \cite{albert00smc}.

\subsection*{Degree correlation}

Are people with high-degrees preferentially linked to those of
high-degrees or low-degrees? To see the assortative mixing with
respect to degree \cite{newman2003mpn}, or degree correlation, one often
calculates the averaged nearest-neighbor degree
\[ k^{\text{nn}}(k)=\sum_{k'=0}^\infty p(k'|k)\ , \]
where $p(k'|k)$ is the probability that a randomly chosen edge has a
vertex with degree $k'$ at either end, while at the other end with
degree $k$.

\figref{fig:stat}~(c) shows $k^{\text{nn}}(k)$ as a function of $k$.
We can observe that in the range $10\lesssim k\lesssim 100$ there is a
positive correlation. Nevertheless, the positive correlation does not
extend to the region $k\gtrsim 100$, where it is slightly negative
instead. This fact can be interpreted in the way that hubs with
high-degrees, say a few hundreds, have propensity to acknowledge
a proposed friendship from anyone who is necessarily in the majority of
lower-degrees. Vertices with degrees of dozens, on the other hand,
tend to form assortative mixing among them as the region of positive
correlation implies. The negative correlation in extremely low-degree
$k\leq3$ is due to the new comers just invited.

Related quantity is the degree-degree correlation, which is the
Pearson correlation coefficient for degrees of vertices $(j_a,k_a)$ at
either end of a link $a$. That is \cite{newman2003mpn},
\[ r=\frac{M^{-1}\sum_a j_a k_a
  -\left[M^{-1}\sum_a\frac{1}{2}(j_a+k_a)\right]^2}{M^{-1}
  \sum_a\frac{1}{2}(j_a^2+k_a^2)
  -\left[M^{-1}\sum_a\frac{1}{2}(j_a+k_a)\right]^2} \ . \]
We obtain the value $r=0.1215\pm 0.0009$, where the standard error was
calculated by the method in \cite{newman2003mpn}. In terms of this
single measure, the correlation coefficient shows a statistical
significance of positive correlation.

\subsection*{Betweenness}

Social interaction between two non-neighboring persons might depend on
another who is on the paths between the first two. A vertex with
relatively low-degree can possibly play an intermediary role in the
flow and diffusion of information. Betweenness centrality
\cite{freeman1977smc} of vertex $v$ is defined by
\[ b(v)=\frac{1}{2}\sum_{s,t\not=v}\frac{\sigma_{st}(v)}{\sigma_{st}} \]
where $\sigma_{st}$ is the number of shortest-paths between a pair of
vertices $s$ and $t$, and $\sigma_{st}(v)$ is the number of such paths
that go through $v$. The factor of $1/2$ takes into account the fact
all shortest-paths are visited twice.

The distribution $p_b$ for $b(v)$ is depicted in the {\it
 cumulative\/} form, $P(b)\equiv\int_b^\infty db'\,p_b$, 
in \figref{fig:stat}~(d). Similar results were obtained in other
networks, especially the power-law tail \cite{goh2002csf}. In our
case, we have $p_b\propto b^{-2.5}$ in the upper-tail regime.

\begin{figure}[htbp]
  \includegraphics[width=0.8\columnwidth]{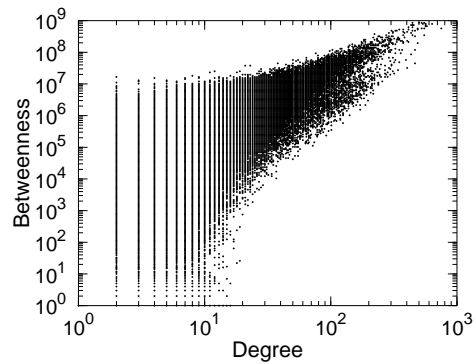}
  \caption{%
    Scatter-plot of degree and betweenness of each vertex.
  }
  \label{fig:deg-bet}
\end{figure}

While the vertices with higher-degrees tend to have higher betweenness
centralities, it is important to see that vertices with relatively
low-degrees have also high betweenness values. We draw the
scatter-plot for the pair of $k$ and $b$ of vertices in
\figref{fig:deg-bet}. While there is obviously positive correlation
between $k$ and $b$, we notice that a same high value of $b$, e.g.
$b=10^6$--$10^7$ in the center of the figure, is produced by vertices
with a wide range of $k$, $20\lesssim k\lesssim 200$. Those vertices
may be connectors between tightly-knit groups of people, and can
provide bridge in the process of acquaintance along friendship.

\subsection*{Number of friends of friends}

Think of your friend's friend who is not your friend. However little
you know about him or her, you might have experienced that your friend
introduced the person to you and that you find that such a person eventually
brings you some new or useful information. The circle of friends of
friends forms a ``horizon'' beyond which you reach to new people and
information. Thus the number of one's friends of friends
gives the size of the horizon.

\begin{figure}[htbp]
  \includegraphics[width=0.8\columnwidth]{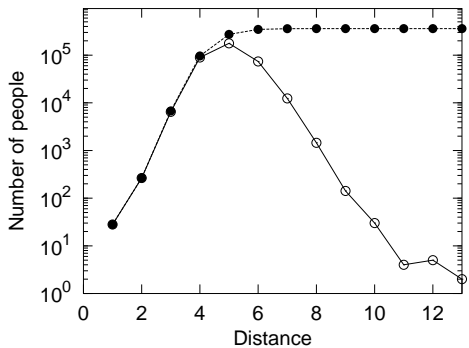}
  \caption{%
    The number of persons who are at distances less than the diameter
    of the network (white circles). The accumulated numbers are shown
    in filled circles.
  }
  \label{fig:horizon}
\end{figure}

\figref{fig:horizon} shows, from a single and typical person, the
numbers of people who are at distance of $d$, where $d\leq
d_{\text{max}}$, and the accumulated numbers at each distance.
Within the distance of six-degree are 96.1\% of people.

In particular, because of the long-tail distribution of degree, the
number of friends of friends is larger than one can
naively expect as $\avg{k}^2$ from the average degree $\avg{k}$. It
may be interesting to compare the average number of friends of
friends in the SNS, denoted by $z_2$, with the value given theoretically in
\cite{newman2001ecn}. The actual value is $z_2=310.6$, while
$\avg{k}^2$ gives 111.5, small by factor of three.

The approximate estimation of $z_2$ with non-vanishing $C_{\text
 g}$ is given by $(1-C_{\text g})(\avg{k^2}-\avg{k})$, which gives the
value 424.0. This is approximation assuming there is no ``squares'',
the case that you know two people who have another friend in common,
but whom you personally do not know. In a further approximation
followed from the assumption that such squares are composed of triangles, one has the
estimate $M_*\,(1-C_{\text g})(\avg{k^2}-\avg{k})$, where
$M_*=\avg{k/[1+C_{\text g}^2(k-1)]}/\avg{k}$. This gives the value
299.4, within 3\% of the actual value.


\section{Community structure}\label{sec:commun}

One feature among the properties of networks which has attracted much
interest is the property of community structure (see
\cite{flake2002soa,radicchi2004dai,palla2005uoc} for example and
\cite{newman2003sfc} for review). Detection of
community structure is to find how vertices in the network cluster
into tightly-knit groups with high density in intra-groups and
with lower connectivity in inter-groups. Without {\it a priori knowledge\/} of
how vertices with similar attributes are assortatively linked to
each other, the community detection would be based solely on the
structure of links.

We use a community-extracting algorithm based on the idea of
modularity introduced by Newman \cite{newman2004fad}. We employ the
implementation developed by Clauset \etal \cite{clauset2004fcs}, which has
made a community-extraction feasible in a practical computational
time for giant networks with millions of vertices (see
\cite{girvan2002css,newman2004fae} for related but different
Girvan-Newman algorithm which is based on edge-betweenness). Let us
call the employed algorithm as the CNM algorithm and the extracted
communities as Newman communities (NCs). Let $e_{ij}$ be
the fraction of edges in the network that connect vertices in group
$i$ to those in group $j$, and let $a_i\equiv\sum_j e_{ij}$. Then
modularity $Q$ is defined by
\[ Q=\sum_i(e_{ii}-a_i^2) \]
which is the fraction of edges that fall within groups, minus the
expected value of the fraction under the hypothesis that edges fall
randomly irrespectively of the community structure.

Detection of community structure is then formulated as an optimization
problem to find a devision of $n$ vertices into mutually disjoint
groups such that the corresponding value of $Q$ is maximum. The
algorithm\cite{newman2004fad} is a greedy optimization algorithm of an
agglomerative hierarchical clustering. The implementation given in
\cite{clauset2004fcs}, when applied to sparse and modular networks,
runs in essentially linear time $\text{O}(n\log^2 n)$.

In each step of the algorithm involves calculating $\Delta Q_{ij}$
that would result from the amalgamation of each pair of groups $i$ and
$j$, choosing the largest of the changes, and doing the corresponding
amalgamation. Because different pairs can give a same amount of
largest change $\Delta Q_{ij}=\Delta Q_{i'j'}$, choice of a particular
pair would alter the subsequent process of amalgamations, resulting in
different community structures as local maxima.

\begin{figure*}[htbp]
  \includegraphics[width=0.8\textwidth]{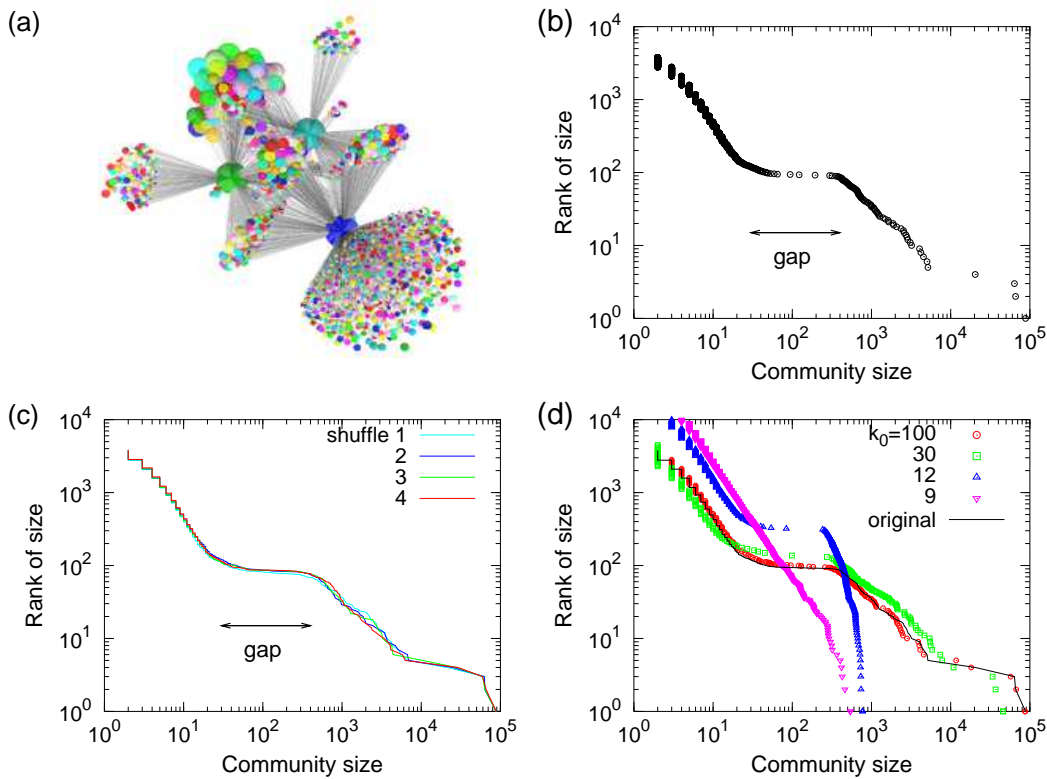}
  \caption{%
    (a)~Visualization of the Newman communities extracted by the CNM
    algorithm \cite{newman2004fad,clauset2004fcs}. Each ball represents a
    community whose logarithmic size is shown in its radius. %
    (b)~The distribution of the community-sizes with its rank in the
    vertical line. The gap shown by a double-ended arrow is the region
    of community-sizes where only few communities are extracted. %
    (c)~Same results as (b) for a few of shuffled storage of edge-lists
    in the network. %
    (d)~Results for a set of subgraphs, which are obtained by deleting
    all vertices with degree $k\geq k_0$ and the edges emanating from
    them. A thick line shows the distribution in (b).
  }
  \label{fig:commun}
\end{figure*}

The output of the algorithm gives the following results. The maximum
modularity is $Q=0.596$, which is considered to be high and to
indicate strong community structure\cite{newman2004fae,newman2004fad}.
Resulting structure includes 3,956 communities. We performed a
coarse-graining visualization by drawing the graph of communities in a
physical model which consists of attractive force between connected
pairs of communities and repulsive force between unconnected pairs.
\figref{fig:commun}~(a) is the visualization, which shows a few large
communities, small-sized numerous communities connected to them, and
medium-sized communities (depicted as a bunch of densely connected
balls in the upper-left portion of the figure),
that are connected mutually as well as to
the large ones. Here size refers to the number of vertices contained
in each community, and is depicted as each ball-size in log scale.
Colors of balls are randomly assigned for the purpose of visibility.

\begin{figure*}[htbp]
  \includegraphics[width=1.0\textwidth]{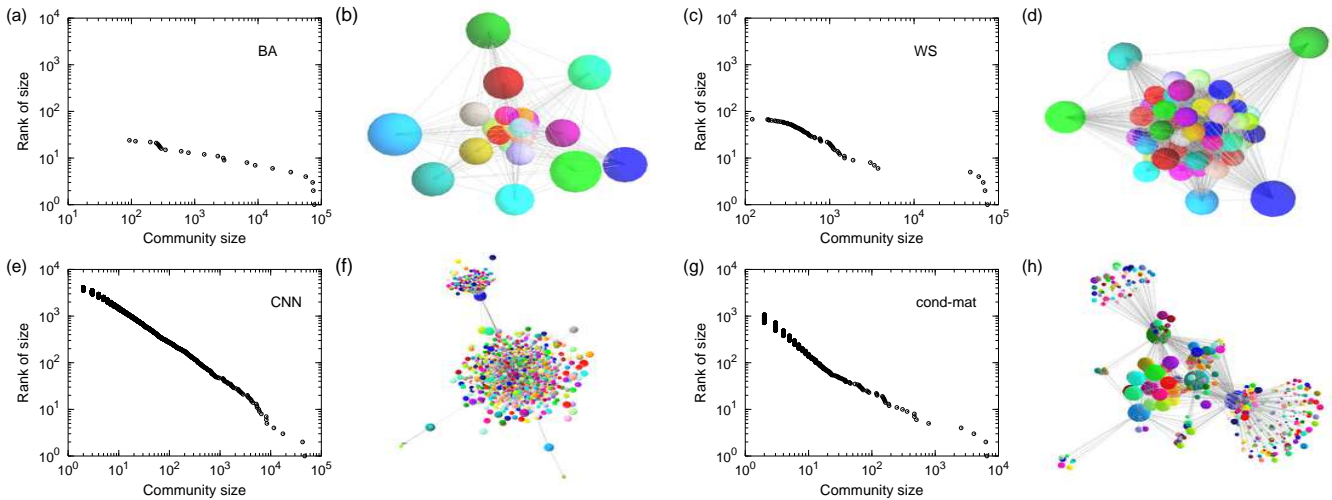}
  \caption{%
    The distributions of community-sizes and visualization of the
    communities for the Barab\'{a}si-Albert model ((a) and (b)),
    the beta model by Watts-Strogatz ((c) and (d)),
    the connecting nearest neighbor by V\'{a}zquez ((e) and (f)), and
    the real-data of collaboration human network in {\tt cond-mat}
    ((g) and (h)).
  }
  \label{fig:commun_diff}
\end{figure*}

The distribution of community-sizes uncovers a novel structure hidden
in the network. \figref{fig:commun}~(b) shows the plot for the
community-size and the rank of the size. In the lower rank
corresponding to the size up to 20, there are numerous small-sized
communities, 3,873 in the number, with 2--20 people in each.  In the
intermediate range of the size between 20 and 400, we found a gap
where few communities are extracted. Up to the size of 4000, there are
80 medium-sized communities with hundreds to thousands people in each
community. Then in the very end of the tail, one sees four largest
communities, whose presence is quite similar to other results of the
CNM algorithm applied for giant networks (see \cite{clauset2004fcs} for
example, and also \figref{fig:commun_diff}~(g)).

Since the algorithm is a greedy optimization as remarked above, one
should check different locally optimal solutions of community
structures. We did so by randomly shuffling the stored order of edge-lists
without changing the network structure, thus effectively altered
the order of amalgamation during the agglomerative clustering.
\figref{fig:commun}~(c) is the rank-size plots for typical outputs,
which shows that the distribution of community-sizes does not differ
for different optimals. Especially, the presence of the gap is
obvious. The value of modularity is estimated as $Q=0.595\pm0.012$,
where the error is the standard deviation for 10 shuffles.

One may expect that the presence of hubs has a considerable effect to
the community structure. It is, however, the case that the vertices of
high-degrees have only a limited effect onto the community structure. In
fact, we take a subgraph consisting of vertices whose degrees are
smaller than a threshold $k_0$, i.e. obtained by deleting the vertices
with $k\geq k_0$ and links emanating from them.
\figref{fig:commun}~(d) shows the results of the CNM
algorithm to these subgraphs.  Even if $k_0$ is
as low as 30, deleting more than 8\% of vertices, the community-size
distribution does not differ significantly.  When $k_0=12$, deleting
25\% of vertices, the gap is still present while exceptionally
large-sized communities are not extracted with this and smaller
thresholds. Only when the threshold is as small as $k_0=9$, one has
many disconnected components with relatively similar sizes, and the
gap disappears. This result implies that it is important to
understand how the majority of vertices with dozens of links are
constructing the overall structure of network.

Previous models of growing networks do not explain the presence of
such a gap found in the distribution of community-sizes. Let us
consider three growth models here. Numbers of vertices and links
in simulations are precisely equal to the numbers of the SNS by
adjusting parameters in each model as follows.
The preferential attachment model, Barab\'{a}si-Albert (BA)
model with each new vertex having degree $m=5,6$ (see the review
\cite{albert00smc}) shows the distribution of community-sizes in
\figref{fig:commun_diff}~(a). For the beta model proposed by
Watts-Strogatz (WS) \cite{watts1998cds} with the rewiring probability
25\%, we have the result in
\figref{fig:commun_diff}~(c). The connecting nearest neighbor (CNN)
proposed by V\'{a}zquez \cite{vazquez2003gnl} with the single
parameter $u=0.81$ (see also below) gives the result in
\figref{fig:commun_diff}~(e).

We summarized some statistical quantities and the resulting
NCs in \tabref{tab:comp} for the models. The numbers of
communities extracted in these models are much smaller than that for
the SNS. Also visualization of the network of communities differs
among the models and from the result for the SNS. The models do not show
a gap in the community-sizes which was observed for the friendship
network of the SNS.

In addition, we performed the CNM algorithm to a
real data of collaboration network in physics community, taken from
{\tt cond-mat}, with $N=30,561$ and $M=125,959$ \cite{palla2005uoc}.
The result is shown in \figref{fig:commun_diff}~(g) with its
visualization in \figref{fig:commun_diff}~(h). While the visualization
for the network of communities is visually similar to the SNS shown in
\figref{fig:commun}~(a), there is obviously no gap in the
community-size distribution.

\begin{table}[htbp]
  \caption{\label{tab:comp}%
    Comparison of degree correlation $r$, global clustering
    coefficient $C_{\text g}$, number of Newman communities $N_{\text{NC}}$ 
    extracted by CNM algorithm and value of modularity $Q$, and
    the characteristics of SF (scale-free), HT (high-transitivity), Gap
    (in the distribution of community-sizes),
    for the real data and the models (see text for details). The models
    have the same numbers of vertices and links as those of the real data.
  }

\begin{ruledtabular}
  \begin{tabular}{lrrrrrccc}
    & $r$ & $C_{\text g}$ & $C$ & $N_{\text{NC}}$ & $Q$
    & SF & HT & Gap \\
    \hline
    {\tt mixi} & 
    0.121 & 0.120 & 0.330 & 3,956 & 0.596 & $+$ & $+$ & $+$ \\
    \hline
    BA\footnotemark[1] &
    -0.009 & 0.0 & 0.0 & 24 & 0.257 & $+$ & $-$ & $-$ \\
    WS\footnotemark[2] &
    0.222 & 0.362 & 0.373 & 68 & 0.685 & $-$ & $+$ & $-$ \\
    CNN\footnotemark[3] &
    0.1 & 0.08 & 0.398 & 1,062 & 0.694 & $+$ & $+$ & $-$ \\
    CNNR\footnotemark[4] &
    0.124 & 0.083 & 0.346 & 5,032 & 0.591 & $+$ & $+$ & $+$ \\
  \end{tabular}
\end{ruledtabular}
\footnotetext[1]{Barab\'{a}si-Albert model \cite{albert00smc}.}
\footnotetext[2]{Watts-Strogatz model \cite{watts1998cds}.}
\footnotetext[3]{Connecting nearest neighbor model \cite{vazquez2003gnl}.}
\footnotetext[4]{CNN model with random linkage (see Section~\ref{sec:cnnr}).}
\end{table}

The list includes a model which we propose in the next section, called
CNNR, connecting nearest neighbor with random linkage.

\section{Connecting nearest neighbors with random linkage}\label{sec:cnnr}

In order to understand why the community-size distribution has a gap,
let us consider how the friendship network in the SNS is formed by
people. The network has the following features.

(i) New vertices are added to the network all the time. The timescale
on which vertices join is not much longer than the timescale on which
they create and break friendship. This may differ from other social
networks.

(ii) Since there is little cost in maintaining a friendship, much
smaller than real-life, people can easily accumulate links
of friendship. A vertex degree is a {\it stock\/}
variable, so to speak, a quantity integrated in time. The long-tail
distribution of degree observed in \figref{fig:stat}~(a)
is partly due to this fact.

(iii) As in many social networks, high transitivity is an important
feature, a process of triadic closure --- people choose new
acquaintances who are friends of friends. The SNS facilitates this
process with various devices as described in Section~\ref{sec:sns}.

(iv) The local clustering coefficient has dependence on the degree as
$C(k)\sim k^{-0.6}$. Additionally, the averaged nearest-neighbor
degree $k^{\text{nn}}(k)$ shows positive degree-correlation in an
intermediate range of degrees, while there is a slight negative
correlation for high-degrees.

Previous studies including \cite{davidsen2002esw,vazquez2003gnl}
suggest that a process of {\it connecting nearest neighbors\/} in a
growth model of network can provide explanation of the features
(i)--(iv). In particular, the concept of {\it potential edge\/}
proposed by V\'{a}zquez \cite{vazquez2003gnl} has a good
interpretation here. A pair of vertices is connected by a potential
edge if they are not connected by a link and they have one or more
common neighbor. Actually, in the context of SNS, people have frequent
occasions to get acquainted with friends of friends by potential
edges.

Unfortunately, however, the community structure studied in
Section~\ref{sec:commun} revealed a feature which cannot be
explained by previous models of connecting nearest neighbors. In fact,
applying the CNM algorithm to numerically
simulated networks generated by the model in \cite{vazquez2003gnl}, we
found that the distribution of community-size for the CNN model, shown
in \figref{fig:commun_diff}~(e), differs from what we
observed for the actual SNS in \figref{fig:commun}~(b).
We thus seek for explanation of the feature:\\
\ (v) The distribution of community-size has a gap or a discontinuity where few
communities are eventually detected by the CNM algorithm.

In social networks including the SNS, individuals are endowed not only
with links, but with sets of characteristics attributed to them.
Examples are association to particular groups with specific interests
(hobbies, thoughts, jobs etc.), living in geographically near regions,
relation of families and relatives, and so on. One gets acquainted
with other people, because one considers them to share one or more
characteristics with oneself, but they may not be in the circle of the
one's acquaintances before. Thus, in addition to connecting nearest
neighbors, people are reaching beyond each circle of friends by making
access along dimensions of characteristics which are often unexpected from what the
current ties show. This process would appear to be {\it random\/} in
the current structure of network, as we assume here.

We propose a model based on these two process, connecting nearest
neighbors with apparently random linkage, which we refer to as CNNR.
This is a simple extension of CNN \cite{vazquez2003gnl}. The model
starts with a single vertex and no links, and iteratively performs the
following.
\begin{enumerate}
\item With probability $1-u$, add a new vertex in the network, create
  a link from the new vertex to a randomly selected vertex $v$. At the
  same time, create a set of potential edges from the new vertex to all the
  neighbors of $v$.
\item With probability $u$, one of the following two processes is
  performed.
  \begin{enumerate}
  \item With probability $1-r$, convert one potential edge selected
    at random into an edge.
  \item With probability $r$, connect one pair of vertices selected at
    random with an edge.
  \end{enumerate}
\end{enumerate}
While a new vertex joins the network with an additional link at the
rate $u$, an edge is either realized from a potential edge or newly
created by random linkage at the rate $1-u$. Therefore, we have $M/N\simeq
1/(1-u)$. The rate $r$ is the relative frequency of random linkage
compared with that of connecting nearest neighbors. If $r=0$, the
model reduces to CNN.

\begin{figure*}[htbp]
  \includegraphics[width=0.95\textwidth]{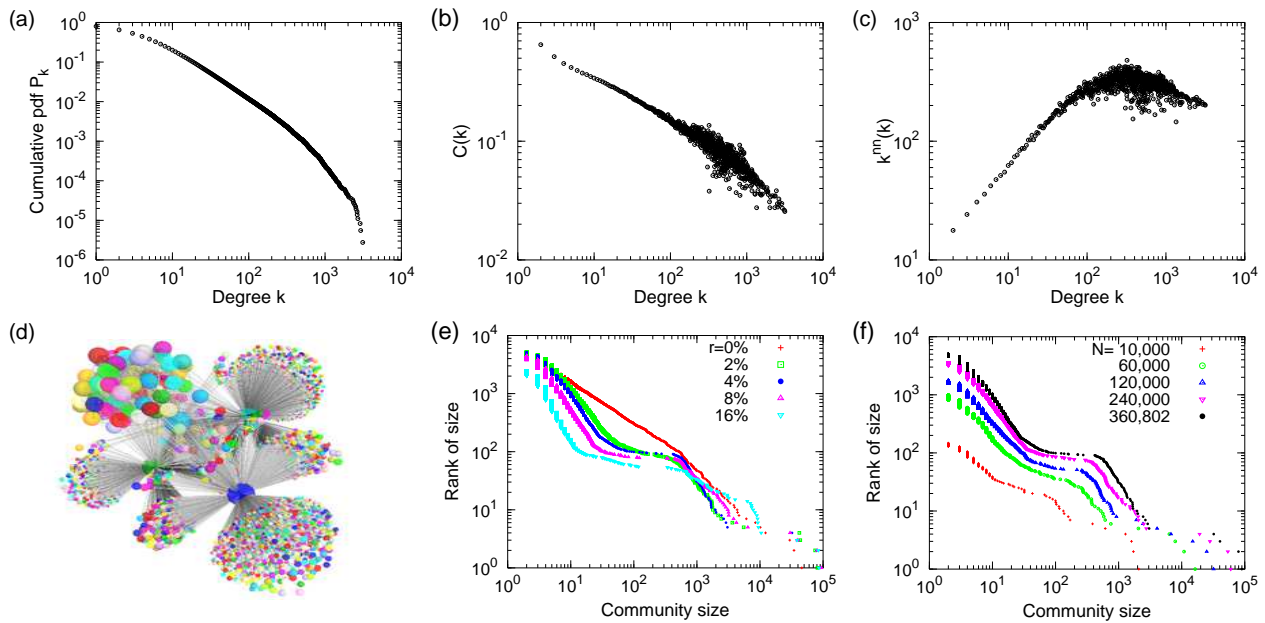}
  \caption{%
    For the model CNNR (connecting nearest neighbor with random
    linkage) with the parameters $u=0.81$ (adjusting $N,M$) and
    $r=0.04$, shown are %
    (a)~the cumulative distribution $P(k)$ for degree $k$, %
    (b)~the local clustering coefficient $C(k)$, %
    (c)~the nearest neighbor degree distribution $k^{\text{nn}}(k)$, %
    (d)~visualization of the communities extracted by the CNM algorithm, %
    (e)~the community-size distributions including the other cases for
    the parameters $r=0.0,0.02,0.08,0.16$, %
    (f)~temporal change of the community-size distribution for
    $r=0.04$ at different sizes $N=1,6,12,24\times10^4$ and 360,802.%
  }
  \label{fig:cnnr}
\end{figure*}

We give a set of results in \figref{fig:cnnr}. The numbers of vertices
and links are adjusted to be equal to $N$ and $M$ for the SNS respectively, by the
parameter $u=0.81$. \figref{fig:cnnr}~(a) is the degree distribution,
having a long tail. \figref{fig:cnnr}~(b) is $C(k)$, which decreases
as $k$ increases. \figref{fig:cnnr}~(c) shows the averaged nearest
neighbor degree $k^{\text{nn}}(k)$, which displays a similar result as
the real-data. These properties are basically the same as the CNN
model \cite{vazquez2003gnl}.

On the other hand, the distribution of community-size has a completely
different shape from \figref{fig:commun_diff}~(e). There exists a gap
in a certain range of community-sizes as shown in
\figref{fig:cnnr}~(e). Note that when $r$ is smaller, the gap is
smaller in its size and vanishes for $r=0$. By comparing the values of
modularity $Q$ for different values of $r$, we suppose that the
parameter $r$ is close to 4\%. Additionally, we can observe in
\figref{fig:cnnr}~(f) that the gap in the distribution of
community-sizes grows larger as the size of the network increases
according to the model of CNNR. We remark that the size of the network
must be large enough in order to detect the presence of the gap.

What does this model tell us about the SNS?
People make the acquaintance of new and yet unfamiliar people more
easily, selectively and inexpensively, far more than what had been
previously possible without such networking sites. But how can one
measure the importance of such augmented acquaintance, in comparison
with other social networks?  Our model could possibly measure
quantitatively the extent with which the apparently random linkage is
at work simultaneously as people enlarge the circle of friends {\it
 via\/} friends of friends. For example, it is our implication that
the process of random linkage takes place much slower in off-line
social networks than it does in the SNS we studied and, quite possibly
in other such social networking sites.
Also one could measure possible difference, among individual
networking sites, of how efficiently the process of random linkage is working
with the help of various designs and devices in social networking sites.

\section{Summary}

We studied the network of mutually acknowledged friendships in the
largest SNS in Japan, currently with more than six million people. In
our dataset when the site is under uniform growth in the access and
in the recruitment, the network is comprised of more than 360,000
people and nearly two million links. By applying to the friendship
network the community-extracting
method developed by Newman and others, we found a novel
feature that there is a certain range of community-sizes for which
only few communities are extracted. This gap in the distribution of
community-sizes was not present in giant human networks such as
co-purchasing data from a large on-line retailer and collaboration
network in physics. Also this is not explained by previous growth
models of networks.

We present a simple model in order to explain this fact as well as
other properties of long-tail degree distribution, correlation
between degree and clustering coefficient, and degree correlation. The
model includes two processes of how people get acquainted with others.
One is connecting nearest neighbors --- acquaintance occurs at
distance of two, friends of friends. And the other represents the
fact that the process of forming links along individual's social
attributes other than the current set of ties, itself, e.g.
to know the presence of persons with same interests,
beyond the circle of friends of friends.

In conclusion, this apparently random linkage is the process that can
explain the gap in the community-size distribution. The two processes
of connecting nearest neighbors and random linkage should be
ubiquitous in social networks, but would be at work with varying
relative frequency. It is our conjecture that the size of the gap will
increase as the network grows further in the SNS. We claim that it
would increase faster than it does in other social networks, as one
could estimate quantitatively based on our model.

\begin{acknowledgments}
  We would like to thank Mixi, Inc. for providing us the
  dataset, in which users are all encrypted except us as three users.
  Our work does not evaluate personality of participants or services
  in any social networking sites. We declare no competing interests.
\end{acknowledgments}

\newpage 

\end{document}